\documentclass{article}
\begin{document}

\centerline{\bf FCC precision requests:}
\centerline{\bf challenges for Monte Carlos and phenomenology tools}
\vskip 2 mm
\centerline{\bf Zbigniew Was}
\vskip 2 mm
\centerline{\it Institute of Nuclear Physics PAN,}

\centerline{  Radzikowskiego 152, Krakow, Poland}
\vskip 5 mm


\abstract{
One can define three levels of ambiguities for the accelerator experiments: worse than 0.3\%, around 0.3\% and better than 0.3\%.
I will address issues important, for the last case, when Monte Carlo programs that simultaneously account for theoretical
and experimental effects are necessary. To complete efforts, simultaneous effort in many specialities over years was necessary.
The monumental projects of Bryan Lynn, Robin Stuart, Dima Bardin, Wolfgang Hollik and contributions of S. Jadach in the domain
of precision physics need to be mentioned, they provide theoretical foundations. Incomplete lists of methodology domains and projects,
usually left into references, appendices, and private notes include: (i) Phase space$\to$ symmetries
(ii) matrix element preparation $\to$ factorizations (iii) program and development process design (iv) testing strategies
(v) user interaction (vi) software tools (vii) partners and competitors.

Usually we were publishing our own projects, let me mention incomplete lists of methodology domains and projects, which were usually
left aside into references, appendices, and private notes: The work started on the basis of previous efforts which can be listed
following names of the programs: names of the programs:
      (i) {\tt FOWL}, (ii) {\tt GENRAP},  (iii) {\tt Mustraal}, (iv) {\tt Koralb},  (v) {\tt Lesko},
      (vi) {\tt Tauola}, (vii) {\tt KoralZ},  (viii) {\tt Lumlog}, (ix) {\tt Oldbab},
      (x) {\tt Bhlumi}, (xi) {\tt Bhwide} and (xii) {\tt KKMC}.
       I will focus on some of these points. Others, hopefully, are covered in other talks.
In particular I do not need to cover exponentiation and some issues of factorization; see contributions to the proceedings by B.F.L.
Ward and A. Tapadar. Developments took years and did not follow a straight line; that is why there are simplifications, biases.
Also a review of literature could not be completed.

\vskip 5 mm
\centerline{\bf IFJ-IV-2026-2}
}

\vfill
\small{17th International Symposium on Radiative Corrections: Applications of Quantum Field Theory to Phenomenology (RADCOR2025)\\
5-10 October 2025\\
Puri, India\\}



\section{Introduction}
One of the aims of high energy physics experiments 
is to confront measurements of interesting observables
with field theory predictions.
Agreements mean \underline{precision tests} of theory and/or measurements
of \underline{coupling constants, masse}s etc. Disagreement points to
\underline{NEW PHYSICS}.
What is needed to perform such a program in the case of accelerator measurements,
especially in cases when precision is essential?
I need to address lifetime  efforts of several people and only as a starting point,
because I want to point to directions for  future work. There are 
  too many aspects to be covered in my talk.

\underline{\bf  Experiments} (even of $pp$ collisions ) best measure directions
of muons, then of electrons, then energies of muons, later of electrons.
Jets are measured less precisely.
From the experimental side, one needs to adopt, acceptance details, often
irregular and direction
dependent. Detectors consist of rectangular (even worse nearly rectangular
and of vague edges)
cells forming a
lattice or better to say kind of 3 dimensional web. In this web
there are dead zones (because of cables). Detector response in barrel and
forward regions differs somewhat.
Sometimes, parts of detectors are faulty, and this makes even more irregular
shapes of acceptance domain.

That, as we have seen in the past, does not bring big complications if precision requirements are not very
high. At 1\% precision level things look fine. The complications arise at
0.3\% precision level. The  methods of lesser precision become complicated,
but are still possible to use. Things change if precision level of 0.1 \% or better is demanded.
These are  essential preconditions from the experimental side,
and they define what is needed from the theory community.
What  are the lessons from the the past.
One may rightfully wonder, why I think 0.3\% ambiguity level defines the border.
The true answer is of course my LEP I experience, later the threshold was not
crossed in a significant manner, but one can expect that its origin is defined by the size of the electromagnetic coupling and detector granularity.
I can not cover details  of the projects, nor provide even partly completed list of publications. Let me refer the  interested reader to references in \cite{Jadach:1993yv,Jadach:1999vf,Jadach:2022mbe,Jadach:1995nk,Janot:2019oyi} for the main projects. Typically each one  involves many man-years of  effort.
LEP I  era measurements~\cite{ALEPH:2010aa} were  often
of precision better than 0.1 \% for  comparison of data with theory predictions
in case of some observables.

First I need to underline importance of efforts not directly
on the phenomenology programs used by the experiments
in theory-data comparisons, but the ones developed earlier. They were essential,
 but more than often their results  needed adaptation and, sometimes painful and complicated, interactions with authors and final users.
  Hopefully that experience  will be useful for work on the future 0.01\%
  precision regime of experiments such as at FCC, where precision at least
\underline{one order of magnitude} better than that at LEP is expected.

Precision predictions require managing {\bf conflicting requirements}.
{\it The most important are those from experiments,} because  massive human
and financial efforts are involved. 
Can phenomenology work ease these burdens? Break no-go limitations?
For high precision or for rare processes where background tails play important role, nothing is easy....

\section{General comments}
Theoretical calculations, because of ultraviolet infinities, must be organized
{keeping in mind renormalization}. That is understandable; that is the way how to
deal with it. That implies that  usually  mass corrections
(thus phase-space details too)  are added
later, sometimes at lower perturbation level.

What does it mean what are the challenges and pressure on solutions,
that must include details from  theory and experiments simultaneously.
To tell the truth, this is not always necessary, for example for
 fitting functions (analytic or semi-analytic) with idealized
acceptance. They  are essential, even now, in the Machine Learning (ML) world.

Intermediate detail level tools 
are helpful, and will remain so in the future.  I was advocating  such solutions
too~\cite{Jadach:1993sa}
(but  in the end, at 0.1\%  this failed). 
\\
{\bf Advantage:} For simplified acceptance with dressed leptons, etc., mass corrections
are less important and, in general, higher order results are easier to get and incorporate.
\\
{\bf Disadvantage:} Detector acceptance details have to be calculated/simulated separately.

I am advocating necessity of the approach where all details of experimental acceptance and theory
predictions can be
evaluated together. But
we have also used  incomplete phase space solutions which worked well and
smoothly
down to 0.5 \% precision level. Let me list them now:
a good example was a
first-order,  matrix element based simulation,  combined with statistically
correlated leading log based simulation with 
collinear photons only: {\tt OLDBAB+LUMLOG}.
The semi analytical approaches like that of Refs.~\cite{Jadach:1990va,Jadach:1988zp} were useful.
But later at 0.3 \%, 0.1\%,
0.043\% levels they gave way to
more refined tools  (but remained as  essential tests \cite{Jadach:1990zf} ).
Another example is 
{\tt KandY}  \cite{Jadach:2001mp}  correlated simulations of 4-fermion final states with initial state 
exponentiation {\tt KORALW} for all tree diagrams and  {\tt YFSWW3} for initial and final state radiation with
 only double resonant diagrams for the hard interaction.
 An  idea is rather a trivial application of the Taylor expansion. Replacing 
 $(1+A+B)^2 \to (1+A)^2+(1+B)^2 -1$  can work well in many conditions.
 One has to define $A$ and $B$ carefully:

 \noindent
A- effects of all tree-level diagrams with respect to double resonant ones
(initial state exclusive exponentiation of QED only)

 \noindent
B- effects of exclusive exponentiation for initial and final state QED effects
but for hard interaction of double resonant diagrams only.

These were also 
 essential steps of work on Monte Carlos, and in many cases enough, even if it brings \underline{massive
difficulties for the experimental users.}

Final, precision solutions were  based on 
simulations with all details of phase space and higher-order induced kinematical configurations 
included, and cutting edge  theoretical effects included.  On the other side, fitting functions
to evaluate impact of theory variations were used as well.
The demands were not easy to follow.
  At LEP I simulation had to take care of configurations \underline{with explicit
  3 bremsstrahlung photons}. For FCC this will probably require 
\underline{5 photons}.
Such configurations are essential to understand effects due to detectors' cell structure, but their inclusion
complicates the implementation of other theory effects.
That points to an important work domain, how to match calculations of distinct perturbation order. 

\noindent
In the future, precision requirements will bring more complications.
One of the necessary steps will include the need to simultaneously include  new processes
like production of   4-fermion, 6-fermion (or jets) final states (of intermediate states like WW, ZZ ZH or tt ).

Many solutions will be working smoothly at 0.5\% level and will also open gate to new applications
as benchmarks.
But finally 
no compromise on phase space details can be assumed.  As {\tt KandY}  was a step forward with respect
to solutions  like  {\tt OLDBAB+LUMLOG}, or {\tt Mustraal+YFS2} installed in earlier version of {\tt KORALZ},
this will have to give way to solutions like {\tt KKMC, Bhlumi} with fitting functions for idealized
but matching simplified kinematics.

In the following let us 
 look at 2-fermion  production like $e^+e^- \to l \bar l$ or
 Bhabha scattering. In these cases  high precision was achieved in comparisons
 between experiments and theory.
There were several steps of the effort.
Separation of SM results into  QED and  hard-interaction effects, effectively of 
genuine weak effects and  vacuum polarization, including some strong interaction loop effects.
Also New Physics
can be often treated as parts of the hard interactions. 
Logically that was the first step, enabling treatment of QED separately.
 For QED,
 eikonal parts of amplitudes for the process of multi-photon production $\beta_0\times \cal{s}....$
 plus corrections where one of $\cal{s}$ in the product was replaced by $\beta_1$ or the
 pair by  $\beta_2$  were necessary to be available
perturbatively. That is the essence of the 
YFS way for  reordering of QED perturbation expansion. Already at the zero-th order
kinematical configurations
with arbitrary large number of photons and all over the phase space become available.
Gauge invariant sectors: initial-state, final-state and their interference (for Bhabha
upper line, lower line and interference) could be established and treated separately.
Note that 
interference effects (real and virtual) can be added at later steps by re-weighting
(with bounded from above weights).  For the interference, contributions to
$\beta_0, \beta_1, \beta_2$ appear later. 
Instead
now, let us turn to foundation of event generation algorithms.

Part of QED consisting of eikonal terms  for initial and final state
bremsstrahlung  can be solved to all orders,  results represent 
{\bf functional} exponent. That includes algorithm for complete phase space
(number of explicit photons is one of the  generation variables too), which  could be constructed
on the basis of incomplete, but known exactly and analytically, amplitudes.

Why the algorithm  was possible?
It is because of \underline{ conformal symmetry} for  eikonal factors and phase space of massless
photons.
That created the basis of algorithm where
initial state photons, invariant mass of intermediate state and final state
photons could be generated independently. Firstly independently generated  photons could be
constrained to 
phase-space limits using scale symmetry.

One should mention a minor difficulty.  To match into the formulae virtual corrections one has to have an
algorithm to deal
 with photon candidates of energies (after re-scaling)  below the lower limit for generation.
The lower limit of photon energy can be arbitrarily low, but should be
the same for all photons to enable matching with virtual corrections and their infrared singularities.
Fortunately, for  that regime, eikonal parts of
amplitudes are  enough.

Once events are  generated with the algorithm based on eikonal parts,
with re-weighting we can introduce  matrix elements of 
 QED: $\beta_1, \beta_2$.
 That looks simple, but it is not always so. For  $\nu_e$ production and $\beta_2$ contribution,
 one has to add charged Higgs ghost exchange Feynman diagrams. 
Is it still QED when  emissions from charged W's must be taken into account?

Now let us turn our attention to non-QED parts of the amplitudes.
The cancellation of singularities requires QED contributions to be organized before
adding hard interactions in the form of form-factors to be introduced for couplings, otherwise gauge
invariance  would be ruined. That worked fabulously well  at 0.1\% precision level and
Center of Mass System Energies (CMS) below 205 GeV. But already at slightly higher,  212 GeV
CMS energy picture started to change; the contribution from $WW$ boxes increased
significantly, deeper into double resonant regime.
 
   A huge effort on genuine weak corrections was needed, even at one-loop level of LEP I times.
   Lots of discussions  accompanied, on so called  star couplings, and at the end on electroweak
   form factors used to parameterize improved Born and to be introduced into QED spin amplitudes. 
   Analytic and anti-analytic constraints of field theory need to be preserved.
   This was proven to work well  at 0.1\% precision level. No compromises were needed.
   Anti-analytic features, optical theorem, Cutkosky rules, were broken only at
   ${\cal{O}} (\alpha^2)\sim 10^{-4}$. This may  not be enough for the FCC regime.
   One may need to revisit and extend this massive effort on  perturbation expansion
   re-ordering, before Monte Carlo implementation can be completed.

\subsection{From the past, summary...}
Even at LEP precision level,  the eikonal part of QED was needed to at least
   \underline{the third order}. Thanks to  re-ordering of perturbation
expansion  one loop genuine weak effects were sufficient.
For FCC that will probably mean two loop genuine weak plus third order QED
and may mean 5-th order for the eikonal in its part?
We have some hints how to start work for 0.01\% precision level,
see e.g.  \cite{FCC:2025lpp} but the question is how to
keep projects ongoing, with  expertise to be passed to new contributors, and let it be extended.

 One should not forget usefulness of 
  older style solutions like {\tt OLDBAB+LUMLOG} or {\tt KORALZ(YFS2) +Mustraal(FSR)} which
  offered a gateway to the framework of NLL-NNLL picture of some 
  ambiguities evaluation.
   In every case,  observable dependent evaluation of ambiguities
  must be in {\it numbers}. 
  Only that is meaningful for measurement-theory comparisons.
  In fact it was succesful for 0.3\% precision level
  and even somewhat beyond. It became, with precision, rapidly
  very laborious and thus not very convenient (to say the least) for
  the experimental users, who had many other things to keep in mind.

 \subsection{Technicality hint from {\tt KKMC} project.}

   My intention was not to talk about internal elements of {\tt KKMC}
    because this is supposed to be covered in other talks.
    Nonetheless, before I will comment on some requirements for the future, I need to mention one essential step: from vector indices
    to spinor indices.
    {\tt KKMC} works not only on spin amplitudes, to enable reduction of number of
    terms, because interferences can be calculated directly from squares of the sum of
    complex interfering contributions (true  for real emissions).
  Vector objects residing in Feynman diagram (like $\not k$) are
  represented as outer product of spinors. That is because of the Kleiss-Stirling spin amplitude language. That was an achievement of CALKUL collaboration~\cite{Kleiss:1985yh}.
  Lots of repetitions in calculations could be avoided.
  It was advantageous and reduced the size of code for formulas a lot.
  It enabled
    useful tests, also  for partial results.
    One should not ignore that from spin amplitudes density matrices for
    $\tau$-pairs
    can be calculated safely and accurately from first principles,
    in the presence of an arbitrary number of photons. Also spin amplitudes
    level was helpful in the definitions of $\beta_i$ terms for
    Yennie-Frautchi-Suura exponentiation and to exploit it fully.
    But adaptation of the techniques was not automatic.
    One should not forget
    laborous preparations of spin amplitudes, necessary for $\beta_i$ calculations, where
    energy-momentum conservation of fxied order calculations does not hold.
    Care is needed, that unphysical huge terms like the ones proportional to 
    $E/m_e$ will not appear.

  \section{Toward the future}
For precision level $\sim$0.01 \% two loop electroweak corrections will be probably needed. Also
third order QED correction and possibly up to 5th order of dominant eikonal parts of QED, because
that means  configurations of 5, to be generated including detector response, photons.
For that purpose, 
two loops electroweak corrections need to have the form with separated,  QED and remaining genuine weak parts, 
and already at the amplitude levels. Only then, results can be  matched, with QED at
higher, that is, third order.
On the other side,
two loop electroweak corrections will impose new pressure on how, also  necessary for precision,
higher than second order QCD corrections can be introduced.
   This is also true about the part of non perturbative QCD effects taken from
   low energy $e^+e^- \to hadrons$ data.
In contrast to what was the case at LEP I,
   for two electroweak loops, higher order  QED-QCD  need to be added not only for s-channel
   or for t-channel exchanges but for both of them  {\underline simultaneously}.
Then, one may  need to review details of  
  complex masses schemes. Solutions proposed by R. Stuart and later by A. Denner, to preserve
  constraints of optical theorem, Cutkosky rules, need to 
remain valid with the first offending terms at ${\cal{O}}(\alpha^3)$ level.

At 0.01 \%  precision level, corrections due to 4-fermion production (and corresponding virtual
corrections) will be needed in full. All that complicates framework 
for calculation and implementation of  bremsstrahlung amplitudes. In particular, effects of interferences
(which are sensitive to delicate cancellations between virtual and real corrections)
and finally crude level generation will need revisiting.
For that purpose the mathematical language of tangent spaces and CW-complexes may be helpful.
CW-complexes may be helpful to systematize
matching of collinear/soft sub-spaces. That was already used in the case of {\tt  PHOTOS} Monte Carlo,
but, so far, with one loop matrix elements only.

If   s-channel and t-channel exchanges contribute simultaneously in a sizable manner, use of
contact interaction and expansion with respect to contact interaction may be helpful.
The inspiring example   \cite{Was:2004ig} for $\nu_e$ may be  helpful in the future as well.
One should not forget 
questions of tests. {\it It is so easy} to make the code working `nearly correctly'.
New tests may  be needed, because one may need to introduce  many adaptations which may make
automatic algebraic manipulations not straightforward to use.

This may result that many 
of the calculations can not be used as off-shelf segments. That is why  there is a quite broad
spectrum of necessary tests to prepare. Such tests, as in the past,
will  sometimes coincide with phenomenological projects.
Big pressure comes from the need to identify the QED part of complete electroweak calculations.

Lots of new testing techniques, theoretical calculations to provide benchmarks, numerical results to
evaluate reliability of physics parts, will be needed.
That means
long time projects. Also from the software side, using physics input, development
of algorithms and  tests will be needed.
For the  computer software engineer perspective, see PhD thesis of Tomasz Przedzinski  
 \cite{Przedzinski:2020axr,Przedzinski:2019who}.
 We should not forget about 
   { manpower and expertise of all subdomains to survive.}
One needs to keep the following in mind:
  How does one identify requested parts of amplitudes? How does one explore properties of 
   Lorentz group, its  sub-groups and corresponding layers. How does one treat extended theories and their
   symmetries.
   Subtracting at the cross section level was found to be useful  at one loop level,
   but what about  higher orders? How does one then avoid negative weight events? In this, 
   beware of detector granularities and details, which are irrelevant at 1\%, appear at about 0.3\%,
   but are essential at 0.1\% or better precision level. See also this conference proceedings,
   contributions
   by A.Tapdar \cite{Tapadar} and B.F.L. Ward \cite{Ward}.
   
\medskip
\medskip
\centerline{\bf Acknowledgements}
\medskip
This work was supported in part by funds from the Polish National Science Centre, Poland,
grant No. 2023/50/A/ST2/00224

\end{document}